\documentclass[10pt,twocolumn]{article}
\usepackage{epsfig,english}
\usepackage{tj}

\def\L#1{\left#1\bgroup}
\def\R#1{\egroup^{\rule{0mm}{1.5mm}}\right#1}

\title{\bf 
Effective Chiral Meson-Baryon Lagrangian from Path Integral Hadronization of
a Quark-Diquark Model}

\author%
{D.~Ebert\thanks{e-mail: \tt{debert@qft2.physik.hu-berlin.de}}
\mbox{}\kern2mm and 
Th.~Jurke\thanks{e-mail: \tt{jurke@pha2.physik.hu-berlin.de}}	
\\
{\small\emph{Institut f\"ur Physik, Humboldt-Universit\"at zu Berlin,
Invalidenstr.~110, 10115 Berlin, Germany}}}
\date{\today{}
\vskip8cm
{\large
{\bf Abstract}\\[5mm]
\begin{tabular}{p{15cm}}
The approach of path integral hadronization,
applied to a quark-diquark toy model, is used to derive an effective
chiral meson-baryon Lagrangian.
In this approach the Goldberger-Treiman relation, found at the
quark-meson level, is re-established at the composite hadron
level. 
For illustrations, electromagnetic low energy characteristics
of composite nucleons are estimated.
\end{tabular}
\vskip1cm}}

\begin{document}

\maketitle

\mbox{}\newpage
\mbox{}\newpage

\section{Introduction}
It is well-known that low energy properties of me\-sons and baryons
can be suitably described by phe\-no\-me\-no\-logical effective
Lagrangians~\cite{We, CaCWZ, EbV81} which embody the global (chiral)
flavour symmetries of quantum chromodynamics ($SU(2)\times SU(2)$ or
$SU(3)\times SU(3)$).  The appearance of an approximate chiral
symmetry and its dynamical breakdown ensures both the generation of
Goldstone pions as well as the existence of current algebra relations
and low energy theorems.

\noindent Concerning meson physics, the approach of path integral
bosonization applied to QCD-motivated Nam\-bu~- Jona-Lasinio (NJL)-type
models~\cite{NaJ} turned out to be a powerful method for deriving
effective chiral meson Lagrangians~\cite{EbV81, EbV83} (for further
references see the recent review~\cite{EbRV}).

\noindent The treatment of baryons as
relativistic bound systems leads us in the case $N_C=3$ to the concept
of diquarks
which can form together with a quark a colourless baryon.
By applying path integral methods it was in particular possible to
derive Faddeev-type equations determining the spectrum of composite
baryons~\cite{CaRP, Re, EbK}.

\noindent It is further a challenge to employ path integral techniques
in collective fields in order to derive effective chiral
meson-baryon Lagrangians, including the hadronic Goldberger-Treiman
relation, directly from an underlying microscopic quark-diquark
interaction.

\noindent We choose in this letter a simple toy model, containing a
local interaction of quarks with (scalar) diquarks in order to
demonstrate the powerfulness of such a method of ``path integral
hadronization''.

\noindent Finally, this model is used to get some rough insight into
the electromagnetic properties of a composite nucleon.

\section{Effective Chiral Meson Lagrangian and Field Transformations}

In order to fix our notations required for the derivation of the
non-linear meson-baryon Lagrangian it is useful to first recapitulate
some results of the bosonization of the NJL-model~\cite{EbV83, EbRV},
emphasizing non-linear field transformations~\cite{EbV81}. Thus, we
consider the NJL Lagrangian with scalar and pseudoscalar couplings
possessing a global flavour and colour $SU(2)_A\times SU(2)_V\times
SU(3)_C$ symmetry

\begin{eqnarray}
\mathcal{L}_{\rm NJL}&=&\bar{q}\L({\rm
i}\hat\partial-m_0\R)q\nonumber\\
&&+
\frac{G}2\L[(\bar{q}q)^2+(\bar{q}{\rm i}\gamma_5\vec\tau q)^2\R],
\end{eqnarray}

\noindent where $\hat\partial=\partial^\mu\gamma_\mu$, $G$ is a
universal coupling constant of dimension (mass)$^{-2}$, $\vec\tau$ are
the Pauli matrices of the flavour group $SU(2)$ and $m_0$ is an
explicit chiral symmetry-breaking quark mass (summation over repeated
indices is always understood).  Introducing collective meson fields
$\sigma$, $\pi_i$ ($i=1,2,3$) by the Gau{\ss} trick

\begin{eqnarray*}
\lefteqn{
{\rm e}^{{\rm i}\intop{\rm d}^4x\kern1mm
\frac G2[(\bar qq)^2+(\bar q{\rm i}\gamma_5\vec\tau q)^2]}
\equiv
\mathcal N_1\intop\mathcal D\sigma\mathcal D\vec\pi\kern2mm\times}
\nonumber\\
&&{\rm e}^{{\rm i}\intop
{\rm d}^4x\L(-\frac1{2G}(\sigma^2+\vec\pi^2)-\bar q(\sigma+{\rm i}\gamma_5
\vec\tau\vec\pi)q\R)},
\end{eqnarray*}

\noindent one gets a semi-bosonized Lagrangian

\begin{eqnarray}
\mathcal L_{\rm NJL}^{qM}&=&\bar q({\rm
i}\hat\partial- \sigma- {\rm i}\gamma_5
\vec\tau\vec\pi)q\kern2mm\nonumber\\
&&-\frac1{2G}\L((\sigma-m_0)^2+\vec\pi^2\R),
\end{eqnarray}

\noindent where the bare quark mass $m_0$ has been absorbed into the
field~$\sigma$.

\noindent It is convenient to define new scalar and pseudoscalar
fields~$\sigma'$ and~$\vec\Phi$ via the exponential parametrization

\begin{eqnarray*}
\sigma+{\rm i}\gamma_5\vec\tau\vec\pi&=
&\left( m + \sigma^\prime\right) {\rm e}^{-\frac{\rm
i}{F_\pi}\gamma_5\vec\tau\vec\Phi},
\end{eqnarray*}

\noindent ($F_\pi$ is the pion decay constant and the constituent
quark mass~$m\equiv \langle\sigma\rangle_0$ is fixed by the gap
equa\-\mbox{tion~\cite{EbV83}).}

\noindent Performing an appropriate chiral rotation of quark fields,%
\footnote{Wess-Zumino anomalies arising due to this
transformation are discarded.}

\begin{eqnarray*}
q&=&{\rm e}^{\frac{\rm
i}{F_\pi}\gamma_5\frac{\vec\tau}2\vec\Phi}\chi,
\end{eqnarray*}

\noindent provides us a Lagrange density in terms of redefined fields:

\begin{eqnarray}
\lefteqn{\mathcal L_{\rm NJL}^{qM}(\chi,\vec\Phi,\sigma^\prime)=}
\nonumber\\[3mm]
&&-\frac1{2G}(m+\sigma^\prime)^2\nonumber\\
&&+\frac{m+\sigma^\prime}{16\kern.5mmG}m_0\kern1mm
{\rm Tr}_{F,D}\left({\rm e}^{-\frac{\rm i}{F_\pi}
\gamma_5{\vec\tau\vec\Phi}}+
{\rm h.c.}
\right)
\nonumber\\
&&+\bar\chi\left\{{\rm i}\gamma_\mu
\left(\partial^\mu+{\rm e}^{-\frac{\rm
i}{F_\pi}\gamma_5\frac{\vec\tau}2\vec\Phi}\kern1mm 
\partial^\mu\kern1mm{\rm e}^{\frac{\rm
i}{F_\pi}\gamma_5\frac{\vec\tau}2\vec\Phi}\right)\right.
\nonumber\\
&&\kern6mm\left.\phantom{\left[\right.}
-m-{\sigma^\prime}^{\rule{0mm}{2mm}}\right\}\chi,
\label{Dirac}
\end{eqnarray}
where the trace is taken over flavour and Dirac indices.

\noindent Note the non-linear transformation law of the meson field
$\vec\xi\equiv \frac{\vec\Phi}{F_\pi}$ under global chiral
transformations $g={\rm e}^{\frac{\rm i}2\gamma_5{\vec a\vec\tau}}
\kern1mm {\rm e}^{\frac{\rm i}2 {\vec v\vec\tau}} \in
SU(2)_A\times SU(2)_V$~\cite{CaCWZ, EbV81} ($a_i$, $v_i$ are real
parameters)

\begin{eqnarray}
g\cdot {\rm e}^{\frac{\rm i}2\gamma_5{\vec\tau\vec\xi(x)}}
&=&
{\rm e}^{\frac{\rm i}2 \gamma_5{\vec\tau\vec\xi^\prime(x)}}
\kern1mm\cdot\kern1mm h(x),
\label{stern}
\end{eqnarray}
where
\[
h(x)={\rm e}^{\frac{\rm i}2\vec\tau 
\vec u^\prime(\vec\xi(x),g)}
\in SU(2)_{V,\rm{loc}}
\]
is an element of the local vector group.

\noindent For later use it is convenient to rewrite the Dirac operator
of the $\chi$-field in~(\ref{Dirac}) by employing the Cartan
decomposition~\cite{EbV81}

\begin{eqnarray}
\lefteqn{\mbox{}\kern-10mm
{\rm e}^{-\frac{\rm i}2\gamma_5{\vec\tau\vec\xi}}
\kern1mm\partial^\mu\kern1mm
{\rm e}^{\frac{\rm i}2 \gamma_5{\vec\tau\vec\xi}}
\kern2mm=}\nonumber\\
&&{\textstyle\frac{\rm i}2 \kern.5mm
\gamma_5\kern.5mm{\vec\tau} }\kern.5mm
\vec\mathcal A^\mu(\xi)\kern1mm+\kern1mm
{\textstyle\frac{\rm i}2 \kern.5mm
{\vec\tau}}\kern.5mm\vec\mathcal V^\mu(\xi).\label{neun}
\end{eqnarray}

\noindent The fields $\chi$, $\vec\mathcal A_\mu$, $\vec\mathcal
V_\mu$ have the following simple transformation law
under~(\ref{stern}) ($\mathcal V_\mu \equiv
\frac{\vec\tau}{2}\vec\mathcal V_\mu$ {\em etc.})

\begin{eqnarray*}
\chi&\mapsto&\chi'\phantom{_\mu}=h(x)\chi\nonumber\\[2mm]
\mathcal V_\mu
&\mapsto&
\mathcal V'_\mu=h(x)\mathcal V_\mu h^\dagger(x)-
h(x){\rm i}\partial_\mu h^\dagger(x)\\[2mm]
\mathcal A_\mu
&\mapsto&
\mathcal A'_\mu=
h(x)
\mathcal A_\mu h^\dagger(x).\nonumber
\end{eqnarray*}

\noindent As we see $\mathcal V_\mu$ transforms as a gauge field with
respect to $SU(2)_{V,{\rm loc}}$. This allows one to define the
following chiral-covariant derivative of the rotated quark
field~$\chi$

\begin{eqnarray}
D_\mu\chi&=&\left(\partial_\mu+{\rm i}\mathcal V_{\mu}
\right)\chi.\label{elf}
\end{eqnarray}

\noindent Thus, using (\ref{Dirac}), (\ref{neun}) and (\ref{elf}), the
inverse propagator of the $\chi$-field takes the form

\begin{eqnarray}
S^{-1}\kern2mm=\kern2mm
{\rm i}\hat D\kern1mm-\kern1mm
m\kern1mm-\kern1mm
\sigma'\kern1mm-\kern1mm
\hat\mathcal A\gamma_5.\label{prop}
\end{eqnarray}

\noindent Finally, by integrating over the $\chi$-field in the
generating functional of~(\ref{Dirac}) and freezing the
$\sigma'$-field, one arrives at a non-linear pion Lagrangian. Indeed,
performing the loop expansion of the resulting quark determinant
(det~$S^{-1}$) and choosing a gauge-invariant regularisation, we
arrive at a mass-like term of the $\mathcal A_\mu(\xi)$-field
contributing to $\mathcal L_{\rm eff}^M$,\footnote{Note that possibly
arising field strength terms of the $\mathcal V_\mu$- and $\mathcal
A_\mu$-fields are vanishing.}
\begin{eqnarray*}
\mathcal L_{\rm
eff}^{M}&=&
\frac{m^2}{g_{\pi qq}^2}{\rm Tr}_F
\mathcal A_\mu^2\kern1mm+\kern1mm\Delta\mathcal L_{\rm sb}.
\end{eqnarray*}
Here $g_{\pi qq}$ is the induced meson-quark coupling constant and
$\Delta\mathcal L_{\rm sb}=\mathcal O(m_0)$ is the symmetry-breaking
mass term given by the second term in~(\ref{Dirac}). As discussed
in~\cite{EbV81} the Cartan form~$\mathcal A_\mu(\xi)$ is just the
chiral covariant derivative of the $\xi$-field, admitting the
expansion 
\begin{eqnarray*}
\vec\mathcal A_\mu(\xi)\kern2mm\equiv\kern2mm D_\mu\vec\xi
&=&
\partial_\mu\vec\xi\kern1mm+\kern1mm\mathcal O(\xi^3)\\[2mm]
&=&
\displaystyle\frac1{F_\pi}\kern.5mm\partial_\mu\vec\Phi
\kern1mm+\kern1mm\mathcal O(\Phi^3).
\end{eqnarray*}
Thus, using the Goldberger-Treiman relation,
\[
F_\pi=\frac{m}{g_{\pi qq}},
\]
at the quark level~\cite{EbV83}, one obtains the effective chiral
meson Lagrangian
\begin{eqnarray}
\mathcal L_{\rm eff}^M&=&\frac{F_\pi^2}2\kern.5mm
D_\mu\vec\xi\kern.5mmD^\mu\vec\xi
\kern1mm+\kern1mm
\Delta\mathcal L_{\rm sb}.\label{14}
\end{eqnarray}

\section{Hadronization of a  Quark-Di\-quark Toy Model}
In order to derive an effective chiral meson-baryon Lagrangian from a
microscopic quark-diquark model, the semi-bosonized
Lagrangian~(\ref{Dirac}) has to be supplemented by the kinetic part of
the scalar isoscalar diquark~$D$, which here will be treated as an
elementary field, and by a quark-diquark interaction term. Let us
consider a simple toy model defined by the following extended chiral
invariant Lagrangian,
\begin{eqnarray}
\mathcal L^{qMD}&=&\mathcal L^{qM}_{\rm NJL}+
D^\dagger\kern1mm\Delta^{-1}\kern1mm
D+\nonumber\\
&&\tilde G\bar\chi D^\dagger D\chi\label{16}\\[5mm]
\Delta^{-1}&=&-\partial_\mu\partial^\mu
-M_D^2,\label{9}
\end{eqnarray}
with diquark mass~$M_D$ and a local quark-diquark interaction with
coupling 
constant~$\tilde G$. 

\noindent As in the NJL model we are now in the position to introduce
collective baryon fields~$B$ using a Gau{\ss} trick. Here we have
\begin{eqnarray}
\lefteqn{\kern-3mm{\rm e}^{{\rm i}\intop{\rm
d}^4x\kern1mm\tilde G\bar\chi D^\dagger D\chi}\kern2mm=}\nonumber\\
&&\kern-7mm\mathcal N^\prime\intop
\mathcal DB\mathcal D\bar B
\kern1mm{\rm e}^{{\rm i}\intop{\rm
d}^4x\kern1mm\L(-\frac1{\tilde G}\bar BB-\bar\chi D^\dagger
B-\bar BD\chi\R)}.\label{17}
\end{eqnarray}
Let us next perform the ``hadronization'' of the
Lagrangian~(\ref{16}) by integrating step by step over quark and
diquark fields in the respective generating functional:
\begin{eqnarray}
\hskip-4mm
Z&=&\mathcal N_1\intop\mathcal D\sigma'\mathcal D\Phi_i
\mathcal DB\mathcal D\bar B\mathcal DD\mathcal
DD^\dagger\kern2mm\times\nonumber\\
&&
{\rm e}^{{\rm i}\intop{\rm d}^4x\kern1mm[-{\rm
i\kern1mm Tr}_{F,D}{\rm ln}\kern1mmS^{-1}-\frac1{\tilde G}\bar
BB]}\times	\nonumber\\ 
&&\kern-2mm\times\kern1mm
{\rm e}^{{\rm i}\intop\kern-1mm\intop{\rm d}^4x\kern1mm{\rm
d}^4y\kern1mm[D^\dagger(x)(\Delta^{-1}-\bar BSB)_{(x,y)}D(y)]},	
\nonumber\\[5mm]
Z&=&\mathcal N_2\intop\mathcal D\sigma'\mathcal D\Phi_i
\mathcal DB\mathcal D\bar B\kern2mm\times\nonumber\\
&&{\rm exp}\left\{{\rm
i}\intop{\rm d}^4x
\left\{-{\rm i\kern0.5mmTr}_{F,D}{\rm ln}\kern1mmS^{-1}\phantom{\frac1{16}}
\right.\right.
\nonumber\\
&&\left.\left.-\frac1{\tilde G}
\bar BB+
{\rm i\kern0.5mmln}(1-\bar BS\Delta B)_{(x,x)}\right\}\right\},\label{19}
\end{eqnarray}
with the quark and diquark propagators~$S$ and $\Delta$ defined
in~(\ref{prop}) and~(\ref{9}), respectively.

\noindent Expanding the logarithms in power series at the one-loop
level (see Fig.~\ref{reihenentwicklung}) and taking into account only
lowest order derivative terms, one describes both the
generation of kinetic and mass terms for the composite baryon fields
as well as the meson-baryon interaction. This yields the expression
{\fontsize{9pt}{13pt}\selectfont
\begin{eqnarray}
\lefteqn{\mbox{}\kern-2.5cm
\mathcal L_{\rm eff}^{MB}=\intop{\rm d}^4x\intop{\rm d}^4y\kern1mm
\bar B(x)\kern1mm\times}\nonumber\\[2mm]
&&\mbox{}\kern-3.5cm\left[\left(-\frac1{\tilde
G}-Z_1^{-1}\gamma_\mu\frac{\vec\tau}2\vec\mathcal V^\mu(x)-
g_A\gamma_\mu\gamma_5\frac{\vec\tau}2\vec\mathcal
A^\mu(x)\right)\delta^4(x-y)\right.\nonumber\\[2mm]
\left.\kern.5mm-\kern.5mm
\Sigma^{\rule{0mm}{3mm}}(x-y)\right] B(y),\label{119}
\end{eqnarray}}
where the baryon self-energy $\Sigma$ admits in momentum space the
decomposition 
$\Sigma(p)=\hat p\kern.5mm\Sigma_V(p^2)+\Sigma_S(p^2)$,
whose low-momentum expansion will be quoted below; 
$Z_1^{-1}$ and $g_A$ are the vector vertex
renormalization and axial coupling constant defined by suitably
re\-gularized loop integrals.

\ABB{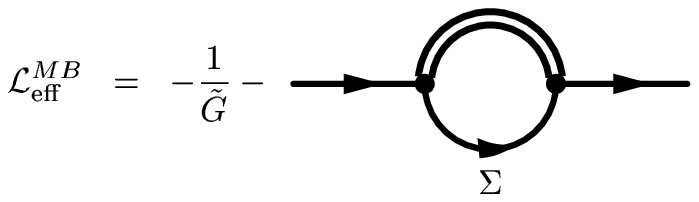}{6.3cm}{}
\mbox{}\vskip-2cm

\ABB{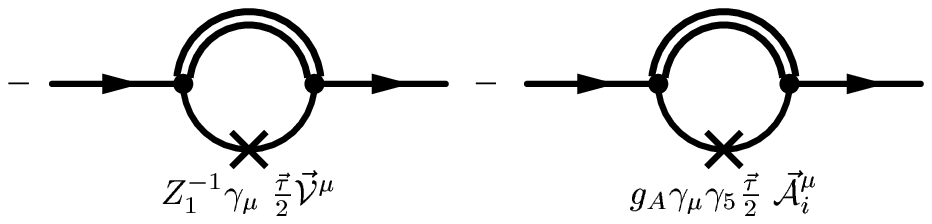}{8cm}{Self-energy contribution of the nucleon,
vector
vertex diagram and axial-vector vertex diagram resulting from
expanding the second logarithm in Eq.~(\ref{19}) of the
text.\label{reihenentwicklung}} 

\noindent Performing the low-momentum expansion of $\Sigma$ and
introducing renormalized fields we have
\begin{eqnarray}
\mathcal L^{MB}_{\rm eff}&=&\bar B_{\rm ren}\L({\rm
i}\hat D-M_B\R)B_{\rm ren}
\nonumber\\
&&-g_A^{\rm ren}\bar B_{\rm ren}\gamma_\mu\gamma_5\frac{\vec\tau}2
B_{\rm ren}\vec\mathcal
A^\mu,\label{ganz}
\end{eqnarray}
with $B=Z^{\frac12}B_{\rm ren}$, $g_A^{\rm ren}=Z\kern.5mmg_A$ and $Z$
being the $Z$-factor derived from the baryon propagator which satisfies
the Ward identity $Z=Z_1$.

\noindent As seen from~(\ref{119}) the nucleon mass is estimated by
\[
\frac1{\tilde G}+\L(M_B\kern.5mm\Sigma_V(M_B^2)+\Sigma_S(M_B^2)\R)=0.
\]

\noindent Recalling $\mathcal A_i^\mu=\partial^\mu\xi_i+\dots$, we
obtained an axial-vector derivative coupling of pions to
baryons.~--~To get rid of the derivative let us redefine the baryon
fields via
\begin{eqnarray*}
B_{\rm ren}&=&
{\rm e}^{-\frac{\rm i}2g_A^{\rm ren}\gamma_5{\vec\tau\vec\xi}}
\tilde B.
\end{eqnarray*}

\noindent Then, the expansion in a power series in~$\vec\Phi$ provides
us the expression
\begin{eqnarray}
\mathcal L^{MB}_{\rm eff}&=&\bar{\tilde B}\L({\rm
i}\kern.5mm\hat\partial-M_B\R)\tilde B\nonumber\\
&&+g_A^{\rm ren}\frac{M_B}{F_\pi}\bar{\tilde B}\kern.5mm
{\rm i}\kern0.5mm
\gamma_5\vec\tau\tilde
B\kern.5mm\vec\Phi+\mathcal O(\Phi^2).\label{fastfertig}
\end{eqnarray}
Obviously we have to identify the constant, associated to the
Yukawa interaction, with the pion-nucleon coupling constant as
\begin{eqnarray}
g_{\bar{\tilde B}\tilde B\Phi}&=&\frac{M_B}{F_\pi}g_A^{\rm ren},
\end{eqnarray}
which is nothing but the Goldberger-Treiman relation at the composite
hadron level. Combining~(\ref{14}) and~(\ref{fastfertig}), the total
meson-baryon Lagrangian reads
\[
\mathcal L_{\rm eff, tot}^{MB}=\mathcal L_{\rm eff}^{M}
+\mathcal L_{\rm eff}^{MB}.
\]

\section{Numerical Discussion}
As a simple application let us calculate the renormalized axial
coupling constant from the matrix element associated to the third 
Feynman diagram of Fig.~\ref{reihenentwicklung},

\begin{eqnarray*}
\lefteqn{{\rm i}M_\mu^{g_A}=}\nonumber\\
&&\mbox{}\kern-5mm-3\kern.3mm{\rm i}\kern.5mm
\intop{\rm i}S(p-k)\kern.5mm\gamma_\mu
\gamma_5 \kern.5mm
{\rm i} S(p-k) \kern.5mm{\rm i} \Delta(k)\kern.3mm\frac{{\rm d}^4k}
{(2\kern.3mm\pi)^4}.
\end{eqnarray*}

\noindent For typical quark and diquark constituent masses (we suppose
an exact isospin symmetry $m_u=m_d=m$) used in such type of approaches

\begin{eqnarray*}
m&=&0.450\rm{\ GeV}\label{erst}\\
M_D&=&0.650\rm{\ GeV}
\end{eqnarray*}

\noindent and a cutoff

\begin{eqnarray*}
\Lambda&=&0.750\rm{\ GeV}\label{letzt}
\end{eqnarray*}

\noindent the model leads to a value

\begin{eqnarray*}
g_A^{\rm ren}=1.07
\end{eqnarray*}

\noindent which is 
in accord with other
related studies (see \emph{e.g.}~\cite{HeW}) but lower then
the experimental value $g_A=1.26$.

\noindent By introducing electromagnetic interactions into the model
it is straightforward to evaluate electromagnetic nucleon form factors
(see Fig.~\ref{bilder}).  

\ABB{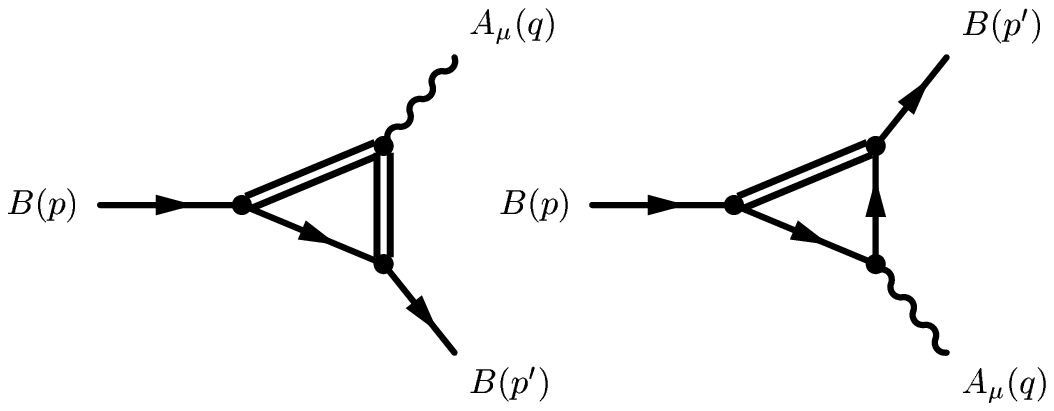}{8cm}{Vertex
diagrams contributing to the electromagnetic form factors of the
nucleon.\label{bilder}}  

\noindent With the above mentioned parameter set
and by taking a diquark form factor ($q=p'-p$),
\begin{eqnarray*}
F_D(q^2)=(1+1.05~{\rm GeV}^{-2}\cdot q^2),
\end{eqnarray*}
suggested in~\cite{WeBAR}, our toy model with scalar diquarks leads to
the following estimates for proton and neutron magnetic moments as well as
electric and magnetic radii, respectively~\cite{Ju}:
\vskip5mm

\begin{center}
{\fontsize{7pt}{10pt}\selectfont
\begin{tabular}{c||r|c||r|c||r|c}
\hline
&$\mu_B$&{\bf exp.}&
$\langle r^2\rangle_e^{\rule{0mm}{4mm}}$&{\bf exp.}&
$\langle r^2\rangle_m$&{\bf exp.}   \\[2mm]
&&&$[$fm$^2]$&{\bf $[$fm$^2]$}
&$[$fm$^2]$&{\bf $[$fm$^2]$}  \\[2mm]
\hline\hline
$P^{\rule{0mm}{4mm}}$&
1.66&{\bf \phantom{$-$}2.79}&0.29&{\bf
\phantom{$-$}0.74}&0.20&{\bf 
0.74}        \\[1mm] 
$N$&$-$0.76&{\bf $-$1.91}&$-$0.01&{\bf $-$0.12}
&0.19&{\bf 0.77}      \\[2mm]
\hline
\end{tabular}}
\end{center}
\vskip-2cm

{\fontsize{9pt}{13pt}\selectfont
\begin{tabular}{rp{5.5cm}}
Table 1:&Estimates within the quark-diquark mo\-del for proton and
neutron magnetic moments as well as electric and magnetic radii.
\end{tabular}
}

\noindent The numerical considerations of the relevant Feynman
diagrams yield only a weak dependence on constituent masses and on the
special choice of the cutoff.

\noindent As a further illustration we want to present here the low
energy dependence of Sachs' electric form factors of the proton and
neutron, respectively (experimental data are taken from~\cite{Ed}).

\vskip-5mm

\ABB{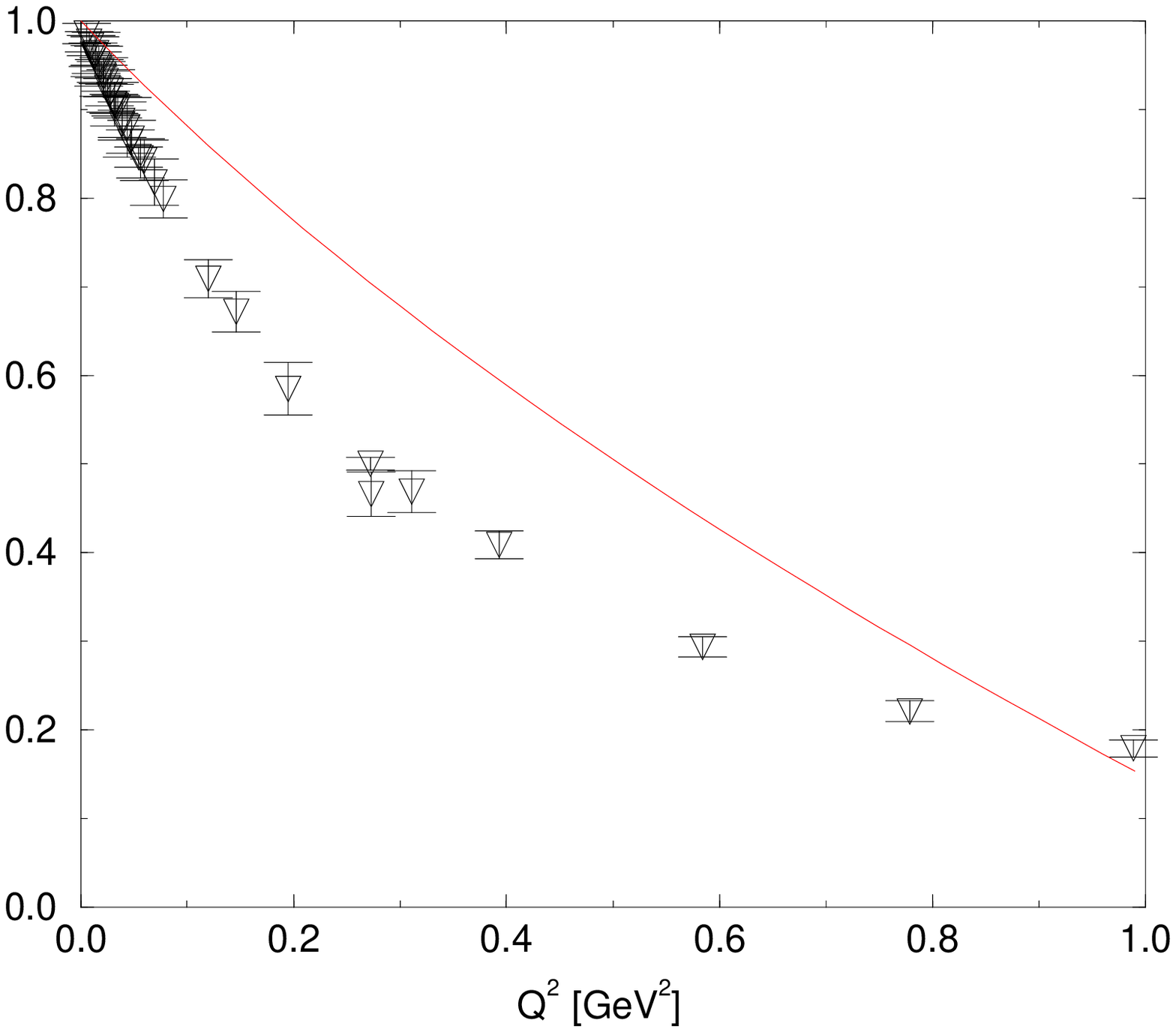}{9.0cm}{Proton electric form factor.}

\vskip-5mm

\ABB{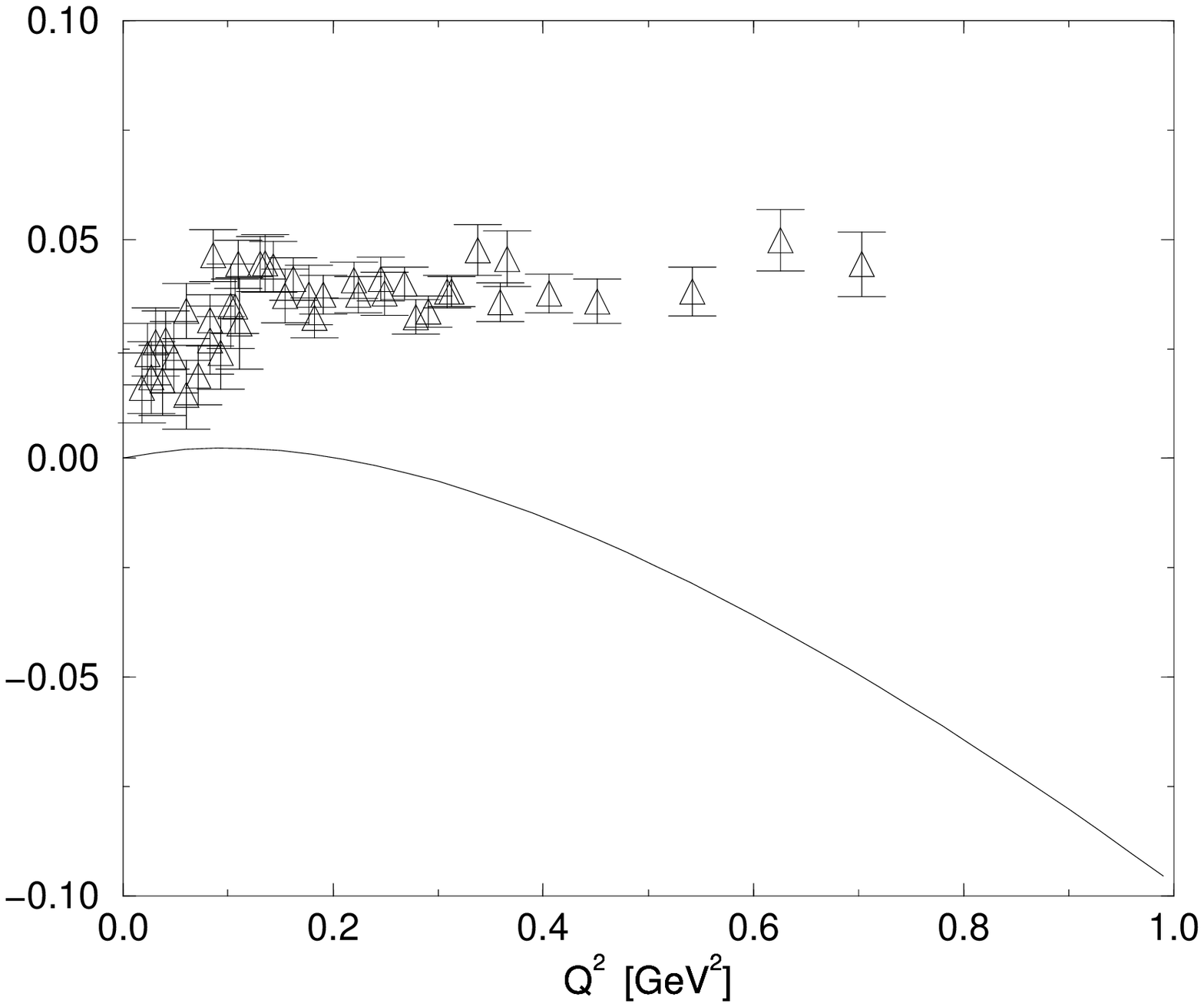}{9.0cm}{Neutron electric form factor.}

\noindent Concerning the estimates shown in the Tab.~1 and Fig.~3,~4, 
it is expected that the observed discrepancies in the
predictions of the toy model from experimental data disappear,
if one uses a more realistic, but also more complicated model
including both non-local quark-diquark
interactions due to quark exchange~\cite{CaRP, Re, EbK} and
axial-vector diquarks.

\section{Summary}
By applying the path integral hadronization method to a simple local
quark-diquark toy model (with scalar diquarks) we have derived an
effective chiral meson-baryon Lagrangian. As an important result the
Goldberger-Treiman relation at the composite meson-baryon level has been
re-established.

\noindent For illustrations, we have further given some numerical
results concerning the axial coupling constant and low-energetic
electric and magnetic form factors of nucleons. A better quantitative
correspondence with data obviously demands the use of non-local
quark-diquark interactions as well as the additional inclusion of the
spin carrying axial diquark.  Especially the large differences in the
magnetic properties (while getting relations $\mu_{\rm prot}/\mu_{\rm
neutr}$ close to experimental findings) are strong hints in this
direction; a statement which is in accord with other recent
papers about this topic~\cite{Ke}.~--~This could be a very promising
subject of further investigations.

\section*{Acknowledgements}
We wish to thank Th.~Feldmann, G.~Hellstern and V.~Keiner for useful
discussions. 

\bibliography{}









\begin{thebibliography}{99}

{\fontsize{9pt}{10pt}\selectfont
\bibitem{We}
S.~Weinberg,
Phys.~Rev.~Lett.~{\bf 18}, 188 (1967).

\bibitem{CaCWZ}
S.~Coleman, J.~Wess and B.~Zumino,
Phys.~Rev. {\bf 177}, 2239 (1969);

C.~Callan, S.~Coleman, J.~Wess and B.~Zumino,
Phys. Rev. {\bf 177}, 2247 (1969).

\bibitem{EbV81}
D.~Ebert and M.K.~Volkov,
Fortschr.~Phys.~{\bf 29}, 35 (1981).

\bibitem{NaJ}  
Y.~Nambu and G.~Jona-Lasinio,
Phys.~Rev.~{\bf 122}, 345 (1961);	

Y.~Nambu and G.~Jona-Lasinio,
Phys.~Rev.~{\bf 124}, 246 (1961).

\bibitem{EbV83}
D.~Ebert and M.K.~Volkov,
Z.~Phys.~{\bf C16}, 205 (1983);

D.~Ebert and H.~Reinhardt,
Nucl.~Phys.~{\bf B271}, 188 (1986).

\bibitem{EbRV}
D.~Ebert, H.~Reinhardt and M.K.~Volkov,
Prog. Part. Nucl. Phys. {\bf 33}, 1 (1994).

\bibitem{CaRP}
R.T.~Cahill, C.D.~Roberts and J.~Praschifka,
Aust. J. Phys. {\bf 42}, 129 (1989).

\bibitem{Re}
H.~Reinhardt,
Phys.~Lett.~{\bf B244}, 316 (1990).

\bibitem{EbK}
D.~Ebert and L.~Kaschluhn,
Phys.~Lett.~{\bf B297}, 367 (1992).

\bibitem{HeW} 
G.~Hellstern and C.~Weiss,
Phys.~Lett.~{\bf B351}, 64 (1995).

\bibitem{WeBAR}
C. {Weiss, A.~Buck, R.~Alkofer and H.~Reinhardt},
Phys. Lett. {\bf B312}, 6 (1993).

\bibitem{Ju}
Th.~Jurke, Diploma thesis, Humboldt-Uni\-ver\-si\-t\"at zu Berlin, 1997.

\bibitem{Ed} 
T.~Eden {\em et al.}, Phys.~Rev.~{\bf C50},
R1749 (1994);

G.~H\"ohler {\em et al.}, Nucl.~Phys.~{\bf B114},
505 (1976);

M.~Meyerhoff {\em et al.}, Phys.~Lett.~{\bf B327},
201 (1994);

S.~Platchkov {\em et al.}, Nucl.~Phys.~{\bf A510},
740 (1990).

\bibitem{Ke}  
V.~Keiner, Z.~Phys.~{\bf A354}, 87 (1996).}

\end{thebibliography}

\endinput
\end{document}